\documentclass{iaus}
\usepackage{graphicx}

\title[Secondary eclipse and TTV of Exo-2b] 
{Searching for the secondary eclipse of CoRoT-Exo-2b and its transit timing variations}

\author[Roi Alonso et al.]   
{Roi Alonso$^1$,
 Suzanne Aigrain$^2$,
 Fr\'ed\'eric Pont$^2$,
 Tsevi Mazeh$^3$,
\and the CoRoT Exoplanet Science Team
}

\affiliation{$^1$Laboratoire d'Astrophysique de Marseille (UMR 6110) 
\\ Technopole de Marseille-Etoile, F-13388 Marseille cedex 13 (France) 
\\ email: {\tt roi.alonso@oamp.fr} \\[\affilskip]
$^2$School of Physics, Univerity of Exeter \\ 
Stocker Road, Exeter EX44QL, United Kingdom \\[\affilskip]
$^3$School of Physics and Astonomy, R. and B. Sackler Faculty of Exact Sciences \\
Tel Aviv University, Tel Aviv 69978, Israel
}

\pubyear{2008}
\volume{253}  
\pagerange{1--6}
\jname{Transiting Planets}
\editors{F. Pont et al., eds.}
\begin{document}

\maketitle

\begin{abstract}
With more than 80 transits observed in the CoRoT light curve with a cadence of 32~s, CoRoT-Exo-2b provides an excellent case to search for the secondary eclipse of the planet, with an expected signal of less than 10$^{-4}$ in relative flux. The activity of the star causes a modulation on the flux that makes the detection of this signal challenging. We describe the technique used to seek for the secondary eclipse, that leads to a tentative 2.5$\sigma$ detection of a 5.5$\times$10$^{-5}$ eclipse. If the effect of the spots are not taken into account, the times of transit centers will also be affected. They could lead to an erroneous detection of periodic transit timing variations of $\sim$20~s and with a 7.45~d period. By measuring the transit central times at different depths of the transit (transit bisectors), we show that there are no such periodic variations in the CoRoT-Exo-2b O-C residuals larger than $\sim$10~s.
\keywords{stars: planetary systems, stars: activity, techniques: image processing, photometric}
\end{abstract}

\firstsection 
\section{Introduction}

\begin{figure}[t]
\begin{center}
 \includegraphics[width=\textwidth]{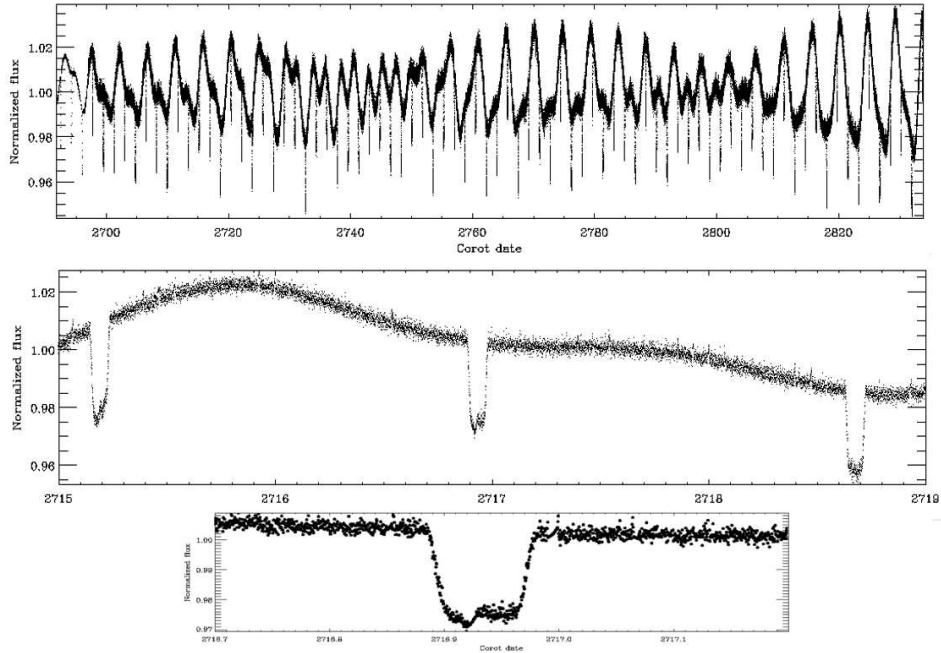} 
 \caption{The CoRoT-Exo-2b light curve (top) and two selected regions showing the effects of the stellar activity, as a modulation of the light curve with the star rotational period (middle), and as alterations of the individual transits resulting from the occultation of stellar spots or active regions (bottom).}
   \label{fig1}
\end{center}
\end{figure}

Launched in December 2006, the CoRoT space mission is producing high quality light curves with outstanding duty cycles of better than 90\% (see the Baglin et al. and Barge et al. papers in this volume). 
Out of the exoplanets detected up to date by this mission, CoRoT-Exo-2b (\cite[Alonso et al., 2008]{alonso08}) is remarkable because of its relatively big size (1.47 R$_J$) and high mass (3.3 M$_J$). Moreover, with its 1.74~d period, CoRoT was able to record more than 80 transits of this planet, in a star that shows clear signs of activity. This activity is recognized because of 1) a $\sim$2\% modulation of the flux of the star with time, caused by the rotation of the star (with a period of $\sim$4.5~d) and evolving star spots, and 2) the effect of star spots being occulted at almost every transit. Both effects are clearly seen in the Fig.~\ref{fig1}. While carrying precious information about the activity of the star, these modulations of the light curve make the search for a secondary eclipse or the measurement of potential transit timing variations (TTV) a challenging task. 

Secondary eclipses in the optical have been searched by different teams, providing upper limits to the albedos (or thermal emission in the infrared) of different exoplanets (e.g., \cite[Charbonneau et al., 1999]{charb99} and  \cite[Collier-Cameron et al. 1999]{collier99} using high S/N spectra, or \cite[Rowe et al. 2008]{rowe08} using the MOST satellite). Both the results of these attempts and the theoretical models (e.g. \cite[L\'opez-Morales \& Seager 2007]{merce07}, \cite[Fortney et al. 2008]{fortney08}) predict a very low contribution of the planetary flux in the optical wavelengths; in the case of pM class planets as defined by \cite[Fortney et al. 2008]{fortney08}, which are characterized by having an appreciable opacity due to Ti0 and VO gases, the planet's thermal emission in the optical is predicted to be more important than the reflected light from the star.

\section{Searching for the secondary eclipse}

If the secondary eclipse were dominated by the reflected light, for this system we would expect an eclipse depth of $\Delta F\sim A \times$6$\cdot$10$^{-4}$, where $A$ is the geometric albedo of the planet. For pM class planets, the geometric albedos in the optical are expected to be very low due to the strong absorption of the incident light by the TiO and VO gases, and the thermal emission of the planet could reach levels of 10$^{-4}$ in the red parts of the visible spectrum (\cite[Fortney et al. 2008]{fortney08}). With an equilibrium temperature of $\sim$1540~K, CoRoT-Exo-2b falls in the pM-class of exoplanets suggested by these authors.

In order to look for this tiny signal, we need to filter the low frequency signal caused by the modulation of the star's flux, while trying not to alter the amplitude of the signal that we are searching.

\subsection{Methodology}

In \cite[Barge et al.(2008)]{barge08a}
and \cite[Alonso et al. (2008)]{alonso08} we described the technique used to compute a phase folded light curve of the transits while minimizing the effect of the low frequency signal from the star. Basically, it consisted in performing low order polynomial fits to the regions around each of the transits, in order to normalize individually each transit. After computing a precise ephemeris, we phase-folded all the normalized transits and binned the flux in phase. The error of each bin was estimated from the dispersion of all the points inside the bin.

To estimate the duration and overall shape of the secondary eclipse, we fitted a trapezoid to the average planetary transit. We expect the secondary eclipse to have the same shape (except for the limb darkening curvature of the transit) as the transit (the radial velocity data indicate negligible orbital eccentricity). In order not to introduce edge effects in our analysis, we treated the light curve from the center of the first transit to the center of the last one, and we removed the transits.

\begin{figure}[t]
\begin{center}
 \includegraphics[width=\textwidth]{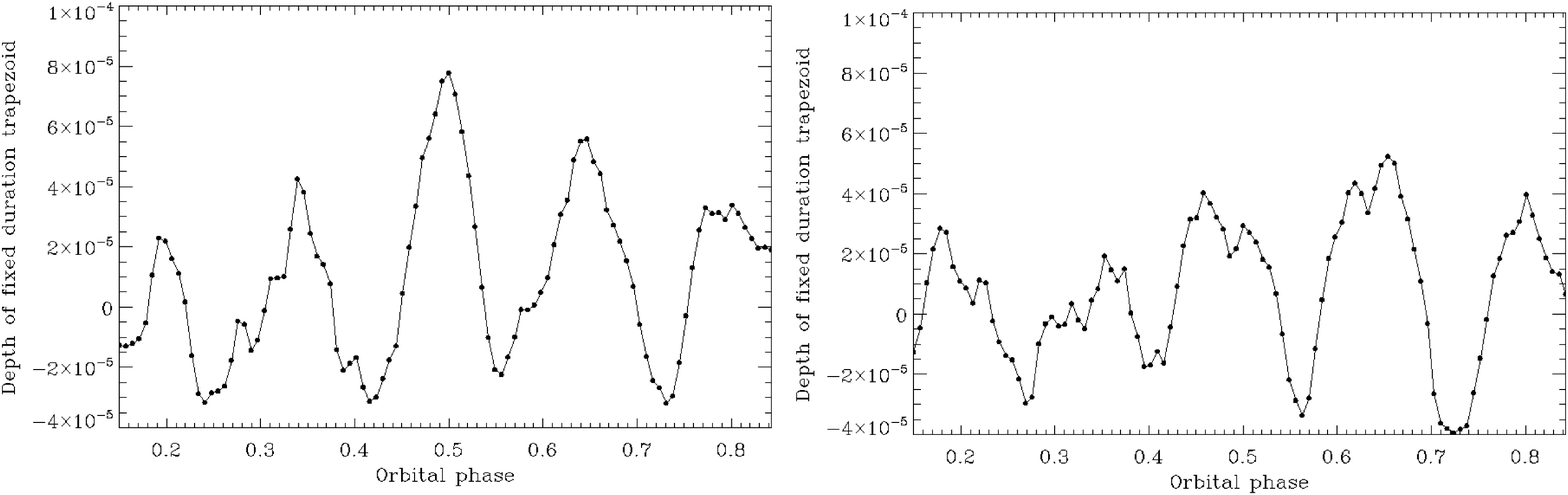} 
 \caption{The depths of fixed duration trapezoids at different orbital phases, for the CoRoT light curve (left) and for a light curve in which the signal from a secondary eclipse has been removed (see text for details). }
   \label{fig2}
\end{center}
\end{figure}

Once the parameters of the expected shape of the secondary are known, we searched for its position and depth. We explored all the orbital phases in which our analysis was not disturbed by the transits, i.e., from orbital phase 0.15 to 0.85. A total of 100 test secondary epochs were tried in this phase range. For each of these test epochs, we proceeded as with the primary transit analysis to filter the stellar activity, then we fitted a trapezoid whose only free parameter was the depth. The result is plotted in Fig.~\ref{fig2} (left). The highest value is well centered on phase 0.5, but some side lobes at phases $\sim$0.33 and $\sim$0.64 are detected, meaning that the procedure is still affected by some residual of the stellar activity. 

\begin{figure}[h]
\begin{center}
 \includegraphics[width=\textwidth]{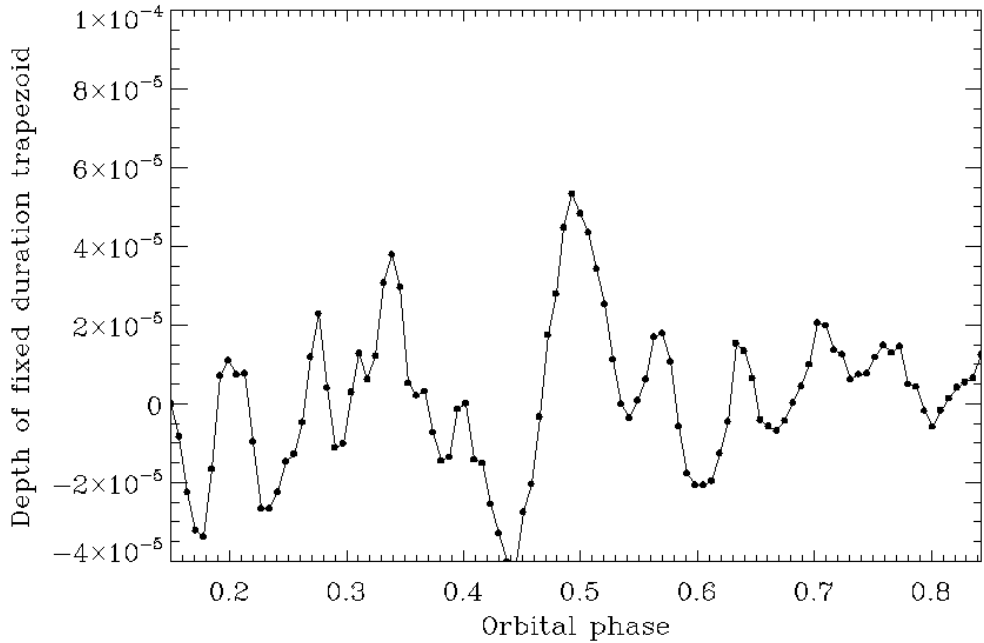} 
 \caption{The difference between the fitted depths of Fig.\ref{fig2} left and right as a function of the orbital phase. The peak close to phase 0.5 represents a tentative 2.5$\sigma$ detection of the secondary of CoRoT-Exo-2b.}
   \label{fig3}
\end{center}
\end{figure}

To evaluate the effect of this residual activity, we constructed a light curve were the potential secondary transits were removed. This was done by using a Savitzky-Golay polynomial filter to model the low frequency signal form the star, removing this signal, shuffling randomly the residuals, thus erasing any potential secondary eclipse, and re-inserting the low frequency signal. With this ``no-secondary" light curve, we repeated the procedure, obtaining the result plotted in Fig~\ref{fig2} (right). The previously detected side lobes are also present in this diagram, meaning that they are an effect of uncorrected low frequency components in the signal. To evaluate the depth and significance of the secondary eclipse detection, we thus subtracted the fitted depths of Fig.~\ref{fig2} (left) to those from Fig.~\ref{fig2} (right). The result, plotted in Fig.~\ref{fig3}, is a diagram less affected by the side lobes. We estimate the significance of the detection from the maximum value of the fitted depth divided by the standard deviation of the fitted depths at all orbital phases, as (5.5$\pm$2)$\times$10$^{-5}$.  As expected for a circular orbit, the highest peak is well centered in phase 0.5.

\section{Looking for transit timing variations}

A total of 81 consecutive transits are present in the CoRoT light curve (Fig.~\ref{fig1}), and the ``alarm mode" switched the time sampling from 512~s to 32~s after the third transit, allowing a better filtering of  the outliers in each of the transits and thus gain in precision on the measurement of each transit central time. 

The TTVs can help to identify unseen additional companions in the planetary system (e.g. \cite[Holman \& Murray 2005]{hol05}, 
\cite[Agol et al. 2005]{agol05}), and several exoplanets have already been searched for TTV (e.g. \cite[Miller-Rizzi et al., this volume]{eri08}). When the star is active, a spurious TTV signal can be created, as we show in the next subsections.

\subsection{A first attempt}

\begin{figure}[t]
\begin{center}
 \includegraphics[width=\textwidth]{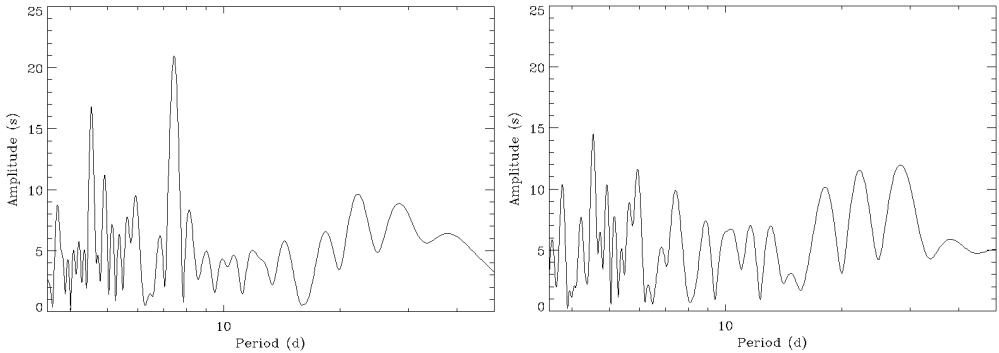} 
 \caption{Fourier amplitude spectra of the O-C residuals of the transit center measurements. Left: using an standard analysis, affected by the activity of the star. Right: Measuring the transit centers from the transit bisectors, as described in the text. The two peaks at the left plot are not present in the right plot, and are an effect of the low-frequency signal from stellar activity.}
   \label{fig4}
\end{center}
\end{figure}

As a part of the process to build a phase folded transit, explained above, we get 81 normalized transits, and the parameters of the best fitted trapezoid to the phase folded transit. We can thus compute the central time of each individual transit by fitting the central time of this trapezoid. The fitting procedure was the IDL implementation of the Levenberg-Marquardt algorithm. The Fourier amplitude spectrum of the observed minus calculated (O-C) residuals is plotted on the left part of Fig.~\ref{fig4}. It shows two apparently significant peaks, one close to the rotational period of the star (4.5~d), and another one at 7.45~d, with an amplitude of $\sim$20~s. While the 4.5~d peak is clearly related to the activity of the star, the 7.45~d peaks appears not related to it, and a fast analysis might erroneously interpret this peak as the effect of a small mass (in the Neptune range) companion. We show below that this peak is also a consequence of the activity of the star, by introducing the concept of ``transit bisectors".

\subsection{Transit bisectors}

\begin{figure}[t]
\begin{center}
 \includegraphics[width=9cm]{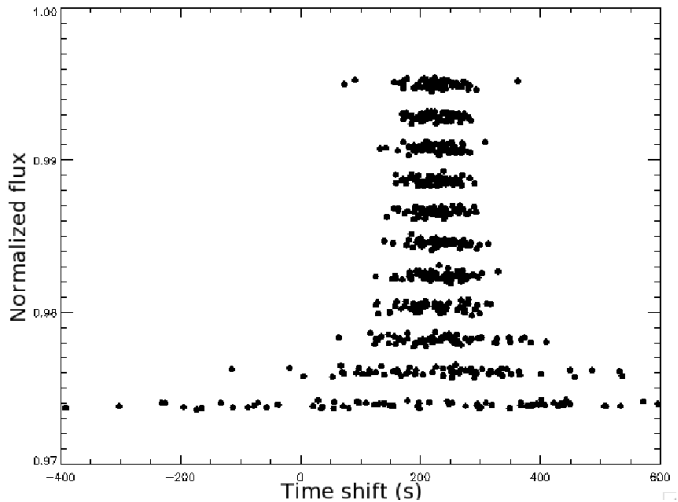} 
 \caption{The transit bisectors of the 81 observed transits. A bigger dispersion of the centers of the transits is seen at the bottom of the transits, as these are the parts more affected by the stellar activity.}
   \label{fig5}
\end{center}
\end{figure}

A transit in which the planet occults active or dark regions of the star can induce spurious shifts of the transit center measurement. In the bottom plot of Fig.~\ref{fig1} for instance, a cross-correlation or a trapezoid fit to the transit will show it appearing earlier than the expected transit time. In order to estimate the transit center, we computed the average times between the data points at 11 different levels of each transit, fitted a low order polynomial to this transit bisector, and extrapolated the fit to the level of the out of eclipse flux. By plotting the bisectors of all the transits together (Fig~\ref{fig5}) we clearly see the effect of a bigger dispersion of the transit centers when measured at the lower parts of the transits as compared to the highest parts. The Fourier amplitude spectrum of the O-C, where the positions were calculated as explained above, now looks as plotted in the right part of Fig.~\ref{fig4}. The tentative peak at 7.45~d now is at the noise level, and we conclude it was induced by the occultation of stellar spots during transits (affecting the bottom of transits more than the sides). We found a possible explanation of this peak as an alias of the rotational frequency of the star ($f_{rot}$): $f_{spur}\sim$2$\cdot$($f_{Ny}-f_{rot}$), where $f_{spur}$ is the frequency of the 7.45~d signal ($f_{spur}$), and $f_{Ny}$ is the Nyquist frequency.

\section{Discussion}

The activity of the host star of CoRoT-Exo-2b is limiting our capabilities to detect the secondary eclipse of this planet, and it can introduce spurious transit timing variation signals. We have made an analysis that shows a tentative 2.5$\sigma$ detection of the secondary eclipse, and employed a new technique to properly measure the individual transit times when the star is active.

In the near future, the light curve of CoRoT-Exo-2b will still be further processed to correct for some residual noises, like the satellite jitter, that might help to improve the significance of the secondary eclipse detection. If the data were dominated by photometric photon noise, and not affected by the stellar activity, the expected S/N ratio for a secondary detection of the depth detected in this paper should be $\sim$5~$\sigma$. Additionally, the photometry of CoRoT-Exo-2b includes information on three different bandpasses (named red, green and blue). As we expect the secondary eclipse to be dominated by the thermal emission from the planet, this signal should be bigger in the red color than in the other two. Even if the correction for the satellite jitter is more difficult in the colored data than in the white light data analyzed in this paper, they might provide further constraints on the depth of the secondary eclipse. We plan to investigate also the behavior of several stars with similar brightness as CoRoT-Exo-2 that were observed with the same aperture shape, in order to search for potential uncorrected systematic noises.

\begin{acknowledgements}
CoRoT is a space project operated by the French Space Agency, CNES, with participation of the Science Programme of ESA, ESTEC/RSSD, Austria, Belgium, Brazil, Germany and Spain.
\end{acknowledgements}

\end{document}